\documentclass[letterpaper]{article}
\usepackage{isea}
\usepackage[pdftex]{graphicx}
\usepackage{times}
\usepackage{helvet}
\usepackage{courier}
\usepackage[final]{changes}
\usepackage{caption}
\usepackage{subcaption}
\usepackage{mciteplus}
\usepackage[numbers]{natbib}

\newcommand{\mycomment}[1]{}

\title{The TSN Building Blocks in Linux}
\author{ Ferenc Fejes\textsuperscript{1}, Péter Antal\textsuperscript{2}, Márton Kerekes\textsuperscript{2}\\
\textsuperscript{1} Ericsson Research TrafficLab, \textsuperscript{2} Budapest University of Technology and Economics\\
Budapest, Hungary\\
ferenc.fejes@ericsson.com\\
\newline
\newline
}
\begin{document} 
\maketitle
\begin{abstract}
Various application areas e.g. industrial automation, professional audio-video, automotive in-vehicle, aerospace on-board, and mobile fronthaul networks require deterministic communication: loss-less forwarding with bounded maximum latency.
There is a lot of ongoing standardization activity in different organizations to provide vendor-agnostic building blocks for Time-Sensitive Networking (TSN), what is aimed as the universal solution for deterministic forwarding in OSI Layer-2 networks.
Furthermore, the implementation of those standards is also happening in Linux. 
Some of them require software changes only, but others have hardware support requirements.
In this paper, we give an overview of the implementation of the main TSN standards in the mainline Linux kernel.
Furthermore, we provide measurement results on key functionality in support of TSN, e.g., scheduled transmission and Linux bridging characteristics.
\end{abstract}

\section{Keywords}

Time-Sensitive Networking, bounded latency, isochron, XDP, ETF, PTP, timestamps, time synchronization, packet delay variation, Ethernet

\section{Introduction}

Some applications have tight Quality of Service requirements in industrial automation, professional audio-video, automotive in-vehicle, aerospace on-board, and mobile fronthaul networks.
These applications cannot tolerate disturbances in their communication.
In everyday web browsing or file download, packet losses in the network are considered part of the normal operation.
These losses are usually handled by the transport layer protocol (e.g. Transport Control Protocol) which retransmits the lost data.
On the contrary, losing a few consecutive packets can be fatal for applications mentioned before.
Note that receiving critical packets after their deadline is equivalent to loss as the application cannot use the information carried in a packet received late.
In the rest of the paper, we refer to those as \textit{time-sensitive applications}, requiring deterministic communication.
Time-sensitive applications require networks to provide the following deterministic characteristics:
\begin{itemize}
	\item High reliability
	\item Bounded maximal latency
	\item Low latency-variation (jitter)
	\item No packet loss due to congestion
\end{itemize}

\textit{Time-Sensitive Networking (TSN)} equip Ethernet networks with tools to meet these requirements.
\mycomment{
}
It is important to note that most applications only require a subset of the TSN tools.
Because of that vendors have been providing different solutions for a long time.
In Layer-2 networks Fieldbus technologies specialized in industrial communication existed since the 1980s.
It would be hard to enumerate all of them, but a few examples: PROFIBUS, Modbus, CC-Link, DeviceNet, etc.
Later, their Ethernet versions also appeared on the market, like EtherNet/IP, PROFINET, EtherCAT, Modbus-TCP, and many others.
Several of them have developed into full-fledged TSN stacks today providing time-sensitive services and APIs at every OSI layer.

\mycomment{
}
Inevitably, a fragmented ecosystem like this has its disadvantages e.g. these solutions are incompatible.
And even if a customer would accept to stick with one vendor and technology, there are industrial equipment simultaneously using multiple fieldbus technologies.
For example, a robotic arm might be connected with multiple network technologies at the same time, one for the control signal, one for safety signaling, and another for diagnostic communication.
Consequently, the manufacturer should support multiple fieldbus ecosystem which increase the development costs and complexity.
Also, in a network there might be devices manufactured only with incompatible fieldbus support.
In that case they can be only used with necessary relays or converters.
At the end, this result high CAPEX and OPEX and low extension or reconfiguration flexibility.

\mycomment{
}

\subsection{IEEE Standards and DetNet}
In response to the growing market demands, the TSN Task Group, as part of the IEEE 802 Working Group \cite{tsntg}, specifies Layer-2 TSN standards.
These standards open the door not only to vendor-independent TSN solutions but also to the implementation of these TSN features in the mainline Linux kernel.
These extend previous standards with TSN capabilities or define standalone TSN standards.
The lower case indicates extensions (802.1Q\textbf{bv}, Q\textbf{ci}, Q\textbf{ch}, etc.), and the upper case indicates stand-alone (802.1\textbf{AB}, \textbf{CB}, \textbf{AS}, etc.).
Standardization of Layer-3 features that are similar to TSN is the subject of the IETF Deterministic Networking (DetNet) Working Group, \cite{detnet} which are not discussed in this paper.

Our paper is divided into two parts.
In the first part, we give a brief overlook of the TSN standard landscape, focusing on their location in the Linux-based ecosystem.
Here we discuss the software and hardware support aspects and the implementation status of the standards as of today.
In the second part, we showcase a few measurements performed on a real testbed.
These experiments give a sense of realistically achievable packet-scheduling precision, the jitter introduced by different switching methods, and also their relation to hardware offload.
Important to note that there are proprietary TSN stacks shipped as binary blobs and precompiled kernel modules in a customized Linux distribution, but these are outside of the scope of this paper.

\section{The TSN Landscape}
\mycomment{

}
A significant advantage of Linux is its support for a wide range of IEEE Layer-2 standards.
This means that incorporating the TSN standards can be accomplished in a well-established framework.
Some TSN capabilities also require explicit support by the hardware.
While for others the hardware support is optional, and the functionality can be implemented purely in software.
The current direction of implementing TSN capabilities in Linux is having the best of both worlds: if there is supporting hardware, the TSN function could be offloaded into that providing very accurate timing precision.
Without TSN-capable hardware, the function can fall back to software mode, so the applications can still leverage them (but with reduced accuracy or increased CPU utilization).
Also advantageous is that Linux has its well-known userspace tooling for network configuration like \texttt{ethtool} or the \texttt{ip}, \texttt{tc}, and \texttt{bridge} tools from the \textbf{iproute2} \cite{iproute2} package.
The user can use those to configure TSN familiarly.

Important to note that to build an end-to-end TSN network, both the end-hosts and the switches should support the related standards.
This is another area where  TSN and Linux have synergy.
Specifically, Linux provides an extensible switch model called \textit{Distributed Switch Architecture} (DSA) \cite{dsa}.
With the help of DSA, switch ports are represented as ordinary network interfaces and can be configured just like them.
There is also a driver model called \textit{switchdev} \cite{switchdev} that helps the switches to offload their dataplane into the hardware.
Switches with DSA and switchdev are first-class citizens in Linux and that way TSN switches are too.

Below we will go through the main TSN standards released so far, and summarize the current status of their adoption in the Linux kernel.

\subsection{Time synchronization: IEEE 1588 and IEEE 802.1AS-2020}

The IEEE 1588 or \textit{Precision Time Protocol} (PTP) \cite{ieee1588} is designed to synchronize clocks over the network with sub-microsecond accuracy.
The protocol implementation spread from the NIC hardware and driver and kernel interfaces to a userspace stack.
The de-facto userspace PTP stack is the \textbf{linuxptp} \cite{linuxptp} providing \texttt{ptp4l} and \texttt{phc2sys} applications (among other tools).
By default the clocks (nor the system clock or the NIC clocks referred as \textit{PTP hardware clock} (PHC)) are not synchronized together, that's done by the \texttt{phc2sys}.
The \texttt{ptp4l} implements the protocol operations like selecting the [grand]master and slave clocks, generating sync packets over the network, etc.
The minimum requirement of PTP is software timestamping support in the NIC driver.
However as we will show it later, if the timestamping is done by the NIC's hardware clock, the accuracy of the synchronization can be improved by orders of magnitudes.

The IEEE 802.1AS-2020 \cite{as2020} define the Generalized PTP (gPTP) to extend the applicability of PTP in various usecases.
One of the differences is that gPTP is restricted to Layer-2 transport while PTP can be used over Layer-3 and 4 (UDP).
Also, gPTP defines designated PTP relays and forbids any other (like tunneled or routed) PTP transmission method between not directly connected devices, which was allowed by regular PTP. Other differences can be found in the 7.5 section of the standard \cite{as2020}.

gPTP defines multiple domains, which is required if we want to run multiple instances of other TSN protocols on one device.
In Linux 5.14 the kernel space PTP virtual clock infrastructure \cite{vlock_kern} was introduced and in 5.18 the userspace interface \cite{vlock_uapi} too.
The linuxptp also got virtual clock support \cite{linuxptp_vclock}, however full gPTP support is still under discussion.

\subsection{Frame Preemption: IEEE 802.1Qbu and 802.3br}

\mycomment{
	Intel FP: https://lore.kernel.org/netdev/20220520011538.1098888-1-vinicius.gomes@intel.com/T/#m2d6688f0971220aabf860a2441a0479eb1511ab8
	NXP FP: https://lore.kernel.org/netdev/20220816222920.1952936-1-vladimir.oltean@nxp.com/t/#m21e1fb0b954f2f42e18b6fb18583ac7af9f63677
}

Frame preemption can suspend the ongoing transmission of a frame for the transmission of another, higher priority frame.
When the high priority frame transmitted, the transmission of the original (preempted) frame is continued.
Let's consider the following scenario: there is a 100 Mbps NIC and the MTU is set to 9000 bytes on it.
In an industrial environment devices optimized to very high reliability and low energy consumption, this is not unusual.
In such a device transmitting an MTU-sized frame took 720 µs. In contrast, there are urgent control or safety signals few bytes in size and they have to wait for the transmission of the large frame to finish.
For example, 64 bytes can be transmitted in 5,12 µs.
IEEE 802.3br \cite{3br} defines \textit{preemptable MAC} and \textit{express MAC} (pMAC and eMAC) where eMAC transmission can interrupt the pMAC transmission at any time.
The IEEE 802.1Qbu \cite{preempt} provides the assignment between the queues and the two kinds of MAC and formalize the management and configuration interface for frame preemption.
With these standards, the operator can configure the urgent transmissions to use the eMAC and everything else to use the pMAC.

Devices with frame preemption hardware support are already on the market, but Linux support is not yet implemented.
Two proposals are discussed on the mailing list currently, one from Intel \cite{fp_intel} and one from NXP \cite{fp_nxp}.
The main difference between the two approaches is the first uses \texttt{ethtool} and \texttt{tc} to configure the pMAC and eMAC and the second only relies on \texttt{ethtool}.
The advantage of the \texttt{ethtool}-only approach is the frame preemption can work on NICs with one transmit queue.

\subsection{Frame Replication and Elimination for Reliability: IEEE 802.1CB}

FRER \cite{1cb} is a standalone standard, not an extension of the IEEE 802.1Q.
Unintuitively from its name it also defines a stream identification function.
That is important for proprietary device compliance and offloading, however, the defined stream identification (like matching on MAC addresses, IPs, or VLAN IDs) is already covered by Linux's filtering capabilities (like \texttt{tc match} or \texttt{tc flower}).

Other than the stream identification the standard defines methods to replicate packets on disjoint paths and then on a device close to the listener drop the unnecessary duplicates (perform elimination).
As a result, the stream remains uninterrupted even if the transmission of the frame fails on some path.

Currently, there is no FRER support in the mainline kernel.
Although there are devices on the market capable to run Linux and offering FRER support in their hardware, their configuration can be done with vendor-provided proprietary tools.
One approach to FRER kernel module, hardware offload, and user-space config interface from NXP were sent to the mailing list \cite{nxp_frer} for discussion, however that assumes DSA tagging which separates the host's traffic from the bridged traffic.

\subsection{Per-Stream Filtering and Policing: IEEE 802.1Qci}

The PSFP \cite{psfp} standardizes functions to perform policing on TSN streams.
The way how to do it is by opening and closing gates in front of ingress frames in a schedule defined by the operator.
For a given stream, if its frames are received when its gate opens,  PSFP let the frames into the bridge.
However, if the gate is closed, it will drop them. The operator can define a byte limit for the open gates too, and if let through enough frames then start dropping them.
On a carefully designed TSN network, the operator has good knowledge of how many bytes should trespass on the gate so more than that could be a result of the malfunction of the talker or malicious activity.
With PSFP also possible to do time-aware reprioritizations of the frames.
This is done by the \textit{Internal Priority Value} (IPV) assignment to the frames (which is metadata, so the frame is not modified) set by the gates.
To match the frames of a given stream, PSFP leverage the stream identification function of the 802.1CB.

PSFP support exists since the 5.8 kernel version \cite{psfp_kernel}.
For stream identification, it uses the \texttt{tc} filters (\texttt{flower}, \texttt{ipset}, \texttt{u32}, etc.) and the actual PSFP functionality implemented in the \texttt{tc gate} action.
Hardware PSFP support from a few devices also exists in the mainline, it can be enabled by passing the \texttt{skip\_sw} option to tc (or we can force the software mode with \texttt{skip\_hw} even on capable hardware).

\subsection{Enhancement for Scheduled Traffic: IEEE 802.1Qbv}

With the help of 802.1Qbv \cite{qbv} standard, the device can do scheduled enqueueing and transmission on the egress frames.
Similarly to PSFP the operator can define the schedule for gate opening and closing.
The schedule contains a list of entries and one entry contains a gate mask (bitmask which tells which stream's gate is open or closed) and duration (for how long).
If the gate is closed for a stream, its frames are enqueued until the gate opens (only dropped if its buffer is full).
If open, the frames pass without any interruption or queueing.
That way the operator can design time windows for the TSN streams with known and bounded latencies, and schedule the best-effort traffic to the remaining time.
This is how we can protect the TSN traffic from congestion.

The 802.1Qbv manifestation in the Linux kernel is the \texttt{taprio} (\textbf{T}ime-\textbf{A}ware \textbf{Prio}rity) queueing discipline (qdisc) \cite{qbv_kern} that made its way to the mainline in version 4.20.
The userspace configration done with tc and the parameters mimic the \texttt{mqprio} qdisc but extending that with the schedule and clock definition.
Hardware offloading is also supported on a few NICs and switches, but important to properly sync the system and PHC clock of the \texttt{taprio} configured NIC to avoid sending frames out of their window (in software mode it is using the system clock).

\subsection{Cyclic Queuing and Forwarding: IEEE 802.1Qch}

CQF \cite{qch} is a standard defining hop-by-hop deterministic frame forwarding for TSN streams.
On the CQF-enabled bridge the operator can configure cycles. At even cycles, the CQF collects frames of one stream and drains the queue of the other, and at odd cycles vica-versa.
That way the two TSN stream never interferes therefore congestion loss cannot happen.
Also because of the cycle-time durations known on each device, the end-to-end latency is upper-bounded and can be easily calculated if we know the number of bridges between the TSN talker and listener. 

The standard explicitly states that CQF operation can be achieved with the coordinated configuration of the PSFP and 802.1Qbv.
As a consequence Linux already has CQF support. For that PSFP (\texttt{tc gate}) should be configured at the ingress port with fully open gates, but with a scheduled assignment of different IPVs in each cycle.
At the egress port \texttt{tc taprio} is configured with the same schedule (same cycle durations).
At the same cycle, PSFP assigns IPV \textbf{\#1} to the frames and \texttt{taprio} keeps the gate closed (filling the queue) for IPV \textbf{\#1} frames.
At the next cycle, \texttt{taprio} opens the gate for IPV \textbf{\#1} (drains the queue) and closes it for IPV \textbf{\#2}, while PSFP assigns IPV \textbf{\#2} to the frames.
And then the schedule restarts this loop.
As one can notice, to successfully do that time-synchronization is required between the ingress and egress ports, and if there are multiple bridges between the talker and the listener, each of them should be in sync.

\section{Implementing Time-Sensitive Applications}

\mycomment{
}

So far we detailed the TSN standard adoption in Linux which had great progress during the past few years.
The first generation of Linux-based TSN switches are already available on the market.
However, for wide adoption of TSN end-hosts must support time-sensitive application development.
For that, it is important to have some kind of support to schedule packets with high precision and receive or process them with bounded delay.

Also like most applications, time-sensitive applications should run in the cloud, which require extra care from the ecosystem, like mapping task scheduling priorities to packet priorities in all network and virtualization layers between the talker(s) and the listener(s).
Cloudification of time-sensitive applications however beyond the scope of this paper.
Below we would like to demonstrate that even sub-microsecond timing precision is achievable on commodity hardware.
Our tests focused on delay variation (jitter) because time-sensitive applications do not tolerate that.
Also, traffic engineering on a TSN network relies on bounded latencies, the important parameter is the maximum latency rather than the average or median.

\subsection{Testbed}

The measurements were performed on three identical generic PC equipped with Intel Core i7-7700K CPU, 8Gb DDR4 RAM, a motherboard with Z270M-PLUS chipset, and Ubuntu 22.04 Server GNU/Linux distribution.
Also to keep up with the recent changes we changed the distro's default (5.15) kernel with 5.19-rc6 kernel version.
Each machine is equipped with a 4 ports Intel I225-LM ethernet interface, which has TSN capabilities and even supports the hardware offloading of some functions.

\mycomment{
The configuration is identical to the mainline Ubuntu kernel, but it was necessary to compile it from the source because we also compiled a version of it with real-time patches applied.
In the following, we refer to the real-time patched kernel as \textbf{rt} and the non-patched as \textbf{regular} for short, including the figures.
}

\begin{figure*}[]
	\centering
	\includegraphics[width=0.7\linewidth]{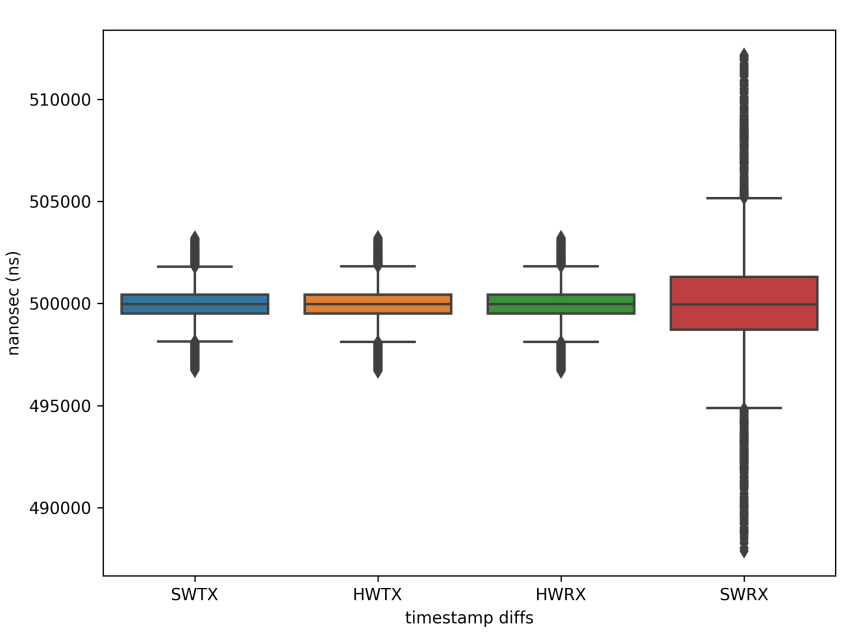}
	\caption[center]{The difference between two timestamps. The packet generator used the sleep method to schedule the transmission of the packets. The (intended) period between two transmission is 500 microseconds}
	\label{fig:aggrtstamps}
\end{figure*}

\subsection{On scheduling precision}

The simplest TSN scenario is where the talker is directly connected with the listener and there are no switches between them.
Intuitively one cannot expect any disturbance in the communication between these two, however, that might not the case.
The talker's application will continuously check the system clock, and send the data when the time is right.
But until the data is copied from the user memory to the kernel space memory, passed down in the network stack by the upper-layer protocol functions to lower-layer ones, copied into the NIC's memory, and finally send to the wire in a frame: that takes time.
Even worse, that time depends on many parameters, like the CPU performance mode, the load generated by other processes, the applied qdisc, and how well the NIC driver is implemented.

To test that in our environment, we generated cyclic traffic between two directly connected machines.
For generating traffic we used the \texttt{isochron} application \cite{iso} which is designed to evaluate TSN switch offload sanity and performance.
The tool is capable to do end-to-end measurements by using the kernel's timestamping infrastructure.
It records 4 timestamps for each packet:
\begin{itemize}
	\item \textit{software tx}: the system clock time saved by the NIC's driver just before the sending started
	\item \textit{hardware tx}: the talker NIC's PHC time when the frame is written into the wire
	\item \textit{hardware rx}: the listener NIC's PHC time when the frame received from the talker
	\item \textit{software rx}: the listener's system clock time saved by the driver right after the packet copied into the kernel memory
\end{itemize}

It's important to keep every clock synchronized, to ensure that \texttt{isochron} does not start the traffic generator until there are high differences between them.
In our scenarios, \texttt{isochron} generated one packet in every 500 µs. To do that accurately, it uses the \texttt{clock\_nanosleep} syscall in absolute time mode provided by the kernel API. After that, it relies on the precision of the kernel's timers to accurately wake up and send the packet. The precision is illustrated in Figure \ref{fig:aggrtstamps}. which summarize the timestamps of 10000 packets.

As one can notice, 80 percent of the software tx timestamps are inside the 1,5 µs radius of the 500 µs.
Then as shown in the figure, the hardware tx and rx timestamps are largely dominated by the precision of the software tx time.
The maximal difference between the intended and the actual timestamps is 3,5 µs in this sample.
However, while the average of the software rx timestamps is identical to the previous ones, 80 percent is in the 5 µs radius and the maximal observed difference is 15 µs.
That is considered fairly precise for most use cases, even taking the time from passing the data received up to the userspace into consideration (a few microseconds additionally).

\begin{figure*}[ht]
	\centering
	\begin{subfigure}[b]{0.45\textwidth}
		\centering
		\includegraphics[width=\textwidth]{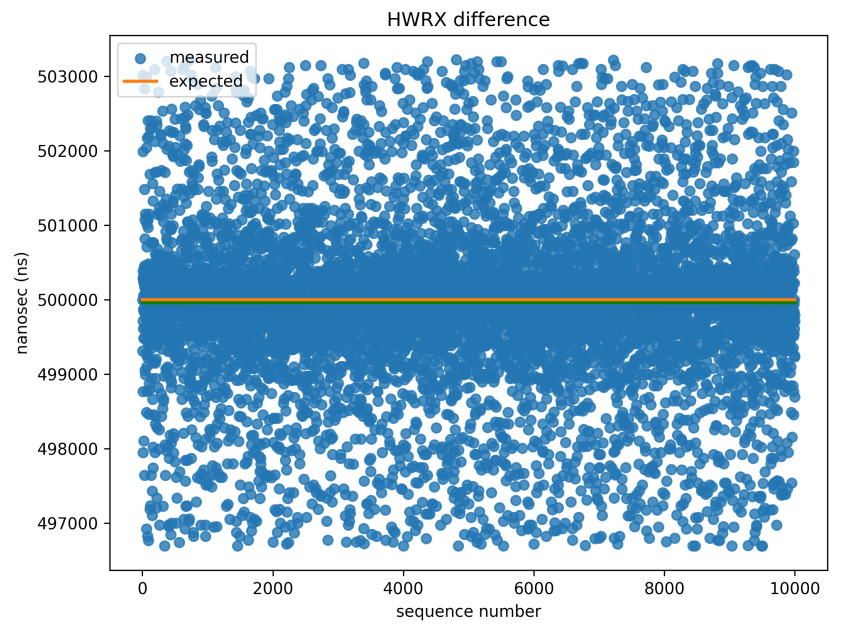}
		\caption{sleep scheduled transmission}
		\label{fig:hwrx_a}
	\end{subfigure}
	\hfill
	\begin{subfigure}[b]{0.45\textwidth}
		\centering
		\includegraphics[width=\textwidth]{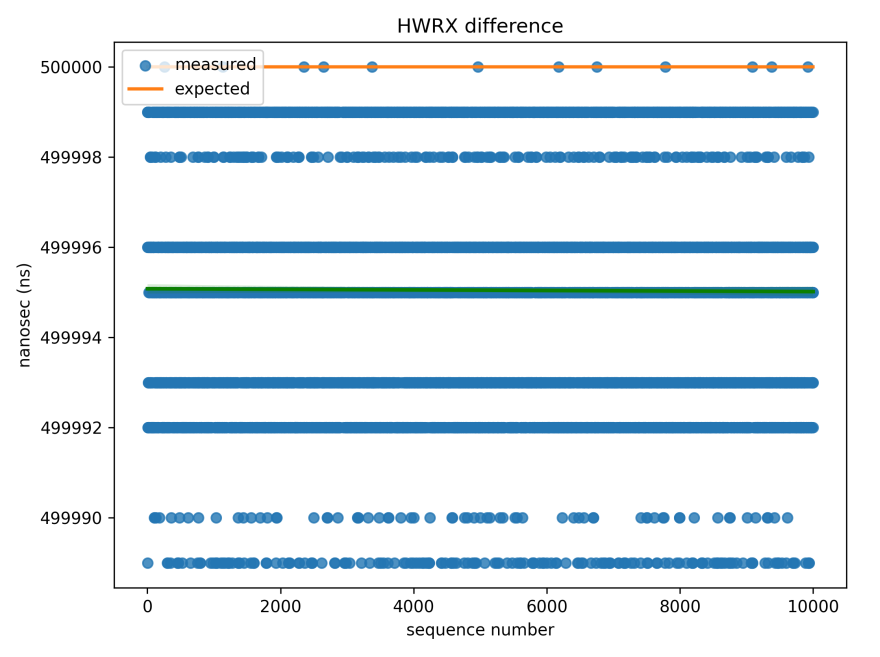}
		\caption{hardware offloaded ETF}
		\label{fig:hwrx_b}
	\end{subfigure}
	\caption{The differences between the receiving timestamps observed by the listener NIC's PHC}
	\label{fig:hwrx}
\end{figure*}

\subsection{SO\_TXTIME and Earliest TX-time First semantic}

For larger packets, the sending syscall together with the copying time might be costly and the jitter can be high even until the data reach the driver.
Also, the \texttt{clock\_nanosleep} precision depends on the system's configuration and performance, which can be less precise on hardware optimized for low energy consumption than in our desktop CPU testbed.

To reduce the jitter, the \texttt{SO\_TXTIME} socket option is introduced. On \texttt{SO\_TXTIME} enabled sockets we can pass a control message (metadata) together with the data containing the timestamp of the expected time of the transmission.
This timestamp is then taken into account by the \textbf{E}arliest \textbf{T}X-time \textbf{F}irst (ETF) qdisc \cite{etf} if applied.
That way the packet is already in the queue of the ETF qdisc (ordered by timestamp) which passes it to the driver if the time arrived.
The advantage of that, ETF qdisc can be offloaded to the NIC hardware.
In that mode, the driver also passes the timestamp to the NIC which schedules the transmission with its PHC.
That way very high precision can be achieved, on a modern NIC that is in the sub-microsecond domain.

To illustrate that, ETF qdisc in offload mode is applied on the talker, and \texttt{isochron} is set to use \texttt{SO\_TXTIME} mode. In figure \ref{fig:hwrx_b}. the whole hardware rx timestamp sample is visualized, and for comparison on figure \ref{fig:hwrx_a}. the previous, \texttt{clock\_nanosleep} method sample is shown. The hardware offload can achieve close to nanosecond precision, the average difference is 5 ns while the maximum observed difference is 11 ns.

\subsection{Jitter introduced by software switching}

Finally, between the talker and the listener, a third PC is inserted to perform software switching.
This is however not a usual TSN scenario because one can expect a hardware switch between the peers.
Nevertheless, it's worth observing how the currently available software switching methods perform because those are very highly customizable and programmable to do other more complex tasks than packet forwarding.
This is essential for the cloudification of TSN functions.

For this test we keep the hardware offloaded ETF packet scheduling to not mistake the talker's jitter with the switching introduced.
Three switching methods examined:
\begin{itemize}
	\item Linux bridge - the Linux kernel's original bridge implementation
	\item XDP - a small eBPF program using the \texttt{bpf\_redirect} function to perform frame redirection between the two interfaces
	\item AF\_XDP - the frame DMA-d into buffers mapped to userspace memory and the redirection done by passing the appropriate descriptors to the TX ring of the other NIC's queue. The tests done by DPDK \texttt{l2fwd} application with AF\_XDP zero-copy backend
\end{itemize}

\begin{figure*}[ht]
	\centering
	\begin{subfigure}[b]{0.3\textwidth}
		\centering
		\includegraphics[width=\textwidth]{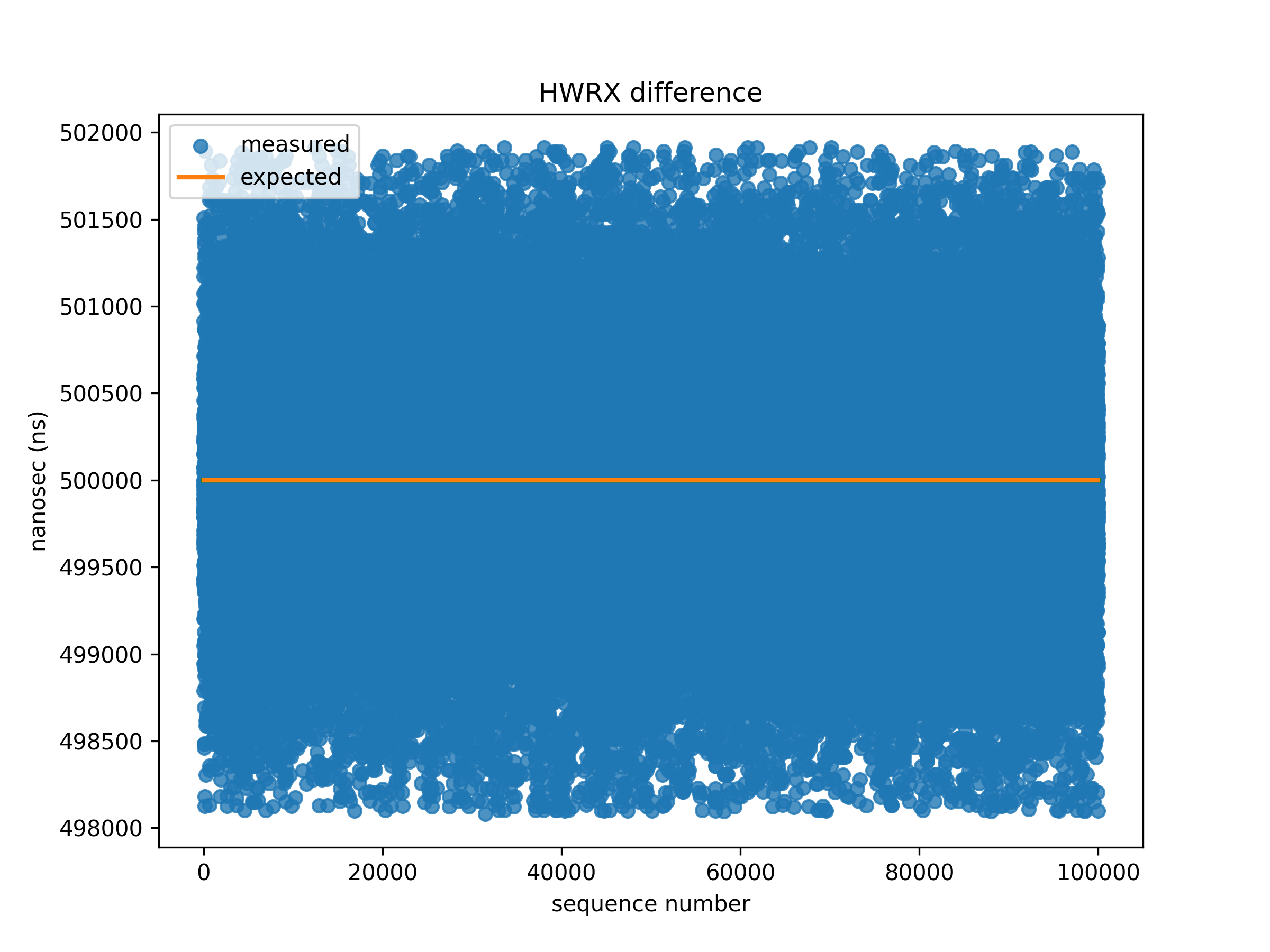}
		\caption{Linux bridge}
		\label{fig:lxbr}
	\end{subfigure}
	\hfill
	\begin{subfigure}[b]{0.3\textwidth}
		\centering
		\includegraphics[width=\textwidth]{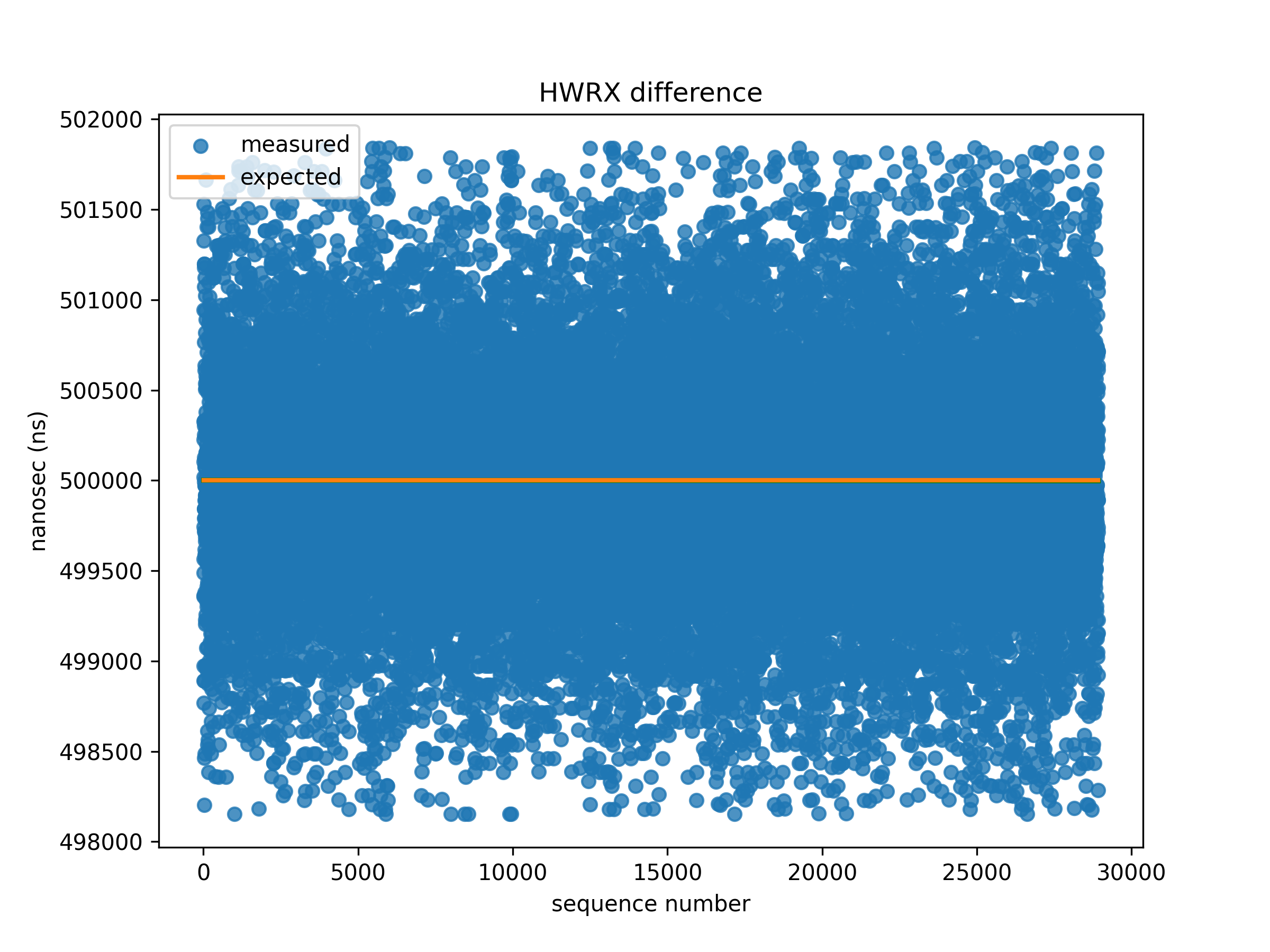}
		\caption{AF\_XDP}
		\label{fig:afxdp}
	\end{subfigure}
	\hfill
	\begin{subfigure}[b]{0.3\textwidth}
		\centering
		\includegraphics[width=\textwidth]{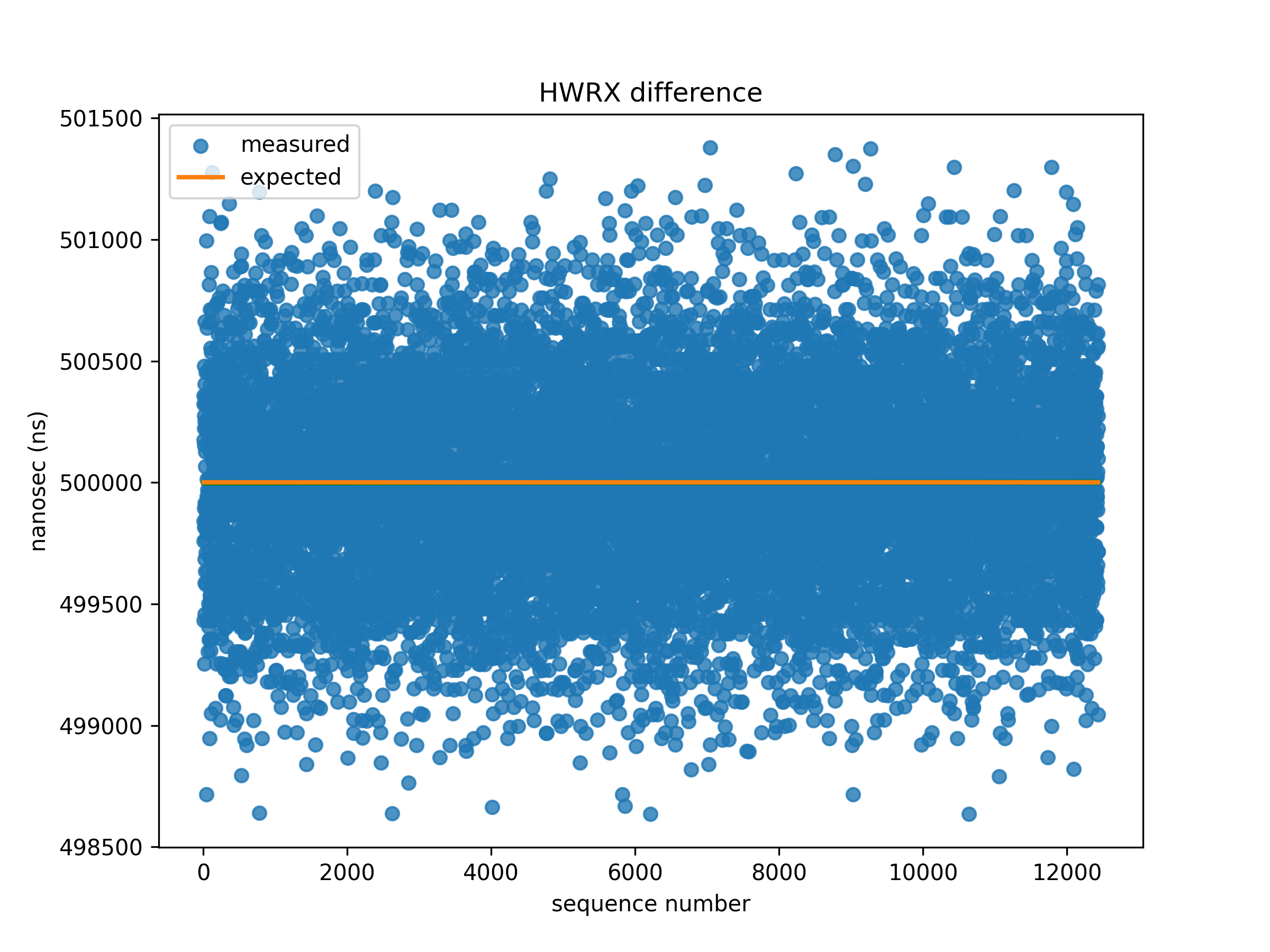}
		\caption{XDP}
		\label{fig:xdp}
	\end{subfigure}
	\caption{The observed jitter at the receiver with different software switching methods}
	\label{fig:switch}
\end{figure*}

During the measurement, CPU and interrupt load on the switch machine are also applied with the stress-ng application.
The conclusion of our testing based on figure \ref{fig:switch}. is there are negligible differences in the jitter introduced by the switching methods.
Linux bridge and AF\_XDP performed equally in terms of maximum observed deviation, and AF\_XDP was a little bit better with the median jitter.
XDP performed the best, because that contains the least complex logic, and just redirects the frame to the egress interface right after its reception.

\mycomment{
	\subsection{Real-Time Linux}
	
	An important aspect of developing time-sensitive applications is to know their behavior on a multiprocess system where they might share the CPU with other applications and the kernel's housekeeping tasks.
	Linux by default is not a hard-realtime kernel, therefore we have no or lose guarantees on deadlines.
	This is especially important when a time-aware application wants to send data periodically with good timing precision. If kernel-introduced jitter (os-jitter for short) causes a high perturbation in the cycle times between the frames actually hitting the wire (copied from userspace to the kernel space memory, then passing through the network stack, DMA-d to the driver), even a well-engineered TSN infrastructure is helpless.
	
	The real-time patches for Linux available for a while. It essentially makes the kernel deterministic with some optimizations notably making soft interrupts preemptable (thread context). Don't mistake determinism with higher performance or lower latency: the real-time kernel doesn't optimize for these, instead, it provides known worst-case latencies.
	In order to investigate the difference between the regular and rt kernel, we used the cyclictest application. This application was developed by the rt kernel developers for performance analysis of the rt kernel. The core of the application is a repeated cycle that sleeps and then computes the difference between the intended and actual sleeping time. Its stores the minimum, maximum, and average latencies. As summarized in the table TODO
}

\section{Conclusion}

\mycomment{
}

In recent years the standardization of TSN triggered vendors to build supporting hardware, both NICs and switches.
Some of these implementations uses mainline Linux kernel, and can gain from its TSN functions.
Adding TSN functions to the kernel usually follows similar steps: first introducing the TSN functionality’s software model with its user
configuration tool, and second if - it’s accepted by the community - hardware offload option is added to the configuration set.
Kernel services support not only the transport node implementations but time-aware applications development as well.
Synchronization is an essential pillar of TSN technology, which resulted in the enhancements of synchronization tools between network devices and the system clock.
Scheduling packets with microsecond precision is made widely available, and with proper hardware support reaching nanosecond level accuracy is also possible.
It is important note, that software-only bridging between the talker and the listener may result in significant jitter.
Furthermore as presented in the measurements section, the listener side jitter is also an important factor for the end2end service.
In longer term the deployment of time-sensitive applications require determinism both at Layer-2 and Layer-3 with appropriate mapping of traffic classes and priorities between them.
As DetNet standardization is evolving, these features may soon be implemented in Linux, built on top of the existing TSN features.
As a summary, the mainline Linux kernel is a performant and flexible platform to rely on for both TSN switch devices and time-aware
applications.

\section{Acknowledgments}

The authors would like to express their appreciation for János Farkas (IEEE 802.1 TSN TG Chair and IETF DetNet Co-chair) and Balázs Varga (IEEE 802.1 TSN TG and IETF DetNet main contributor) for their help in clarifying the terminology and improving the readability of the paper.
They also want to say thanks to Vladimir Oltean and Vinicius-Costa Gomes for their help on the netdev mailing list, quickly responding to our questions and to András Bogár for his valuable comments on the text.

\bibliographystyle{IEEEtranMN}
\renewcommand*{\bibfont}{\normalfont\footnotesize}
\bibliography{isea}

\end{document}